\documentclass[%
 reprint,
 amsmath,amssymb,
 aps,
]{revtex4-1}
\usepackage{physics}
\usepackage{lmodern}
\usepackage{xcolor}
\usepackage{url}
\usepackage{hyperref}
\usepackage{cleveref}
\usepackage{graphicx}
\usepackage{comment}

\usepackage[utf8]{inputenc}
\usepackage[T1]{fontenc}
\hypersetup{
  colorlinks=true,
  citecolor=magenta,
  linkcolor=blue,
  urlcolor=violet
 }
\DeclareUnicodeCharacter{2212}{\ensuremath{-}}

\def\be{\begin{equation}}\def\ee{\end{equation}}
\newcommand{\baa}{\begin{equation}\begin{aligned}}
\newcommand{\ea}{\end{aligned}\end{equation}}

\newcounter{para}
\newcommand\mypara{\par\refstepcounter{para}\textbf{\thepara .}\space}

\newdimen\hfuzz
\newdimen\vfuzz

\begin{document}

\title{\textbf{de Sitter Excited State in Heterotic ${\rm E}_8 \times {\rm E}_8$ Theory}}

\author{Suddhasattwa Brahma$^1$}\email{suddhasattwa.brahma@gmail.com}
\author{Keshav Dasgupta$^2$}\email{keshav@hep.physics.mcgill.ca} 
\author{Bohdan Kulinich$^2$}\email{bohdan.kulinich@mail.mcgill.ca}
\author{Archana Maji$^3$}\email{archana$\_$phy@iitb.ac.in}
\author{P. Ramadevi$^{3}$}\email{ramadevi@iit.ac.in}
\author{Radu Tatar$^4$}\email{Radu.Tatar@liverpool.ac.uk}

\affiliation{$^1$Higgs Center for Theoretical Physics, University of Edinburgh, Edinburgh, EH9 3FD, UK}

\affiliation{$^2$Department of Physics, McGill University, Montr\'eal, Qu\'ebec, H3A 2T8, Canada}

\affiliation{$^3$Department of Physics, Indian Institute of Technology Bombay, Mumbai 400076, India}

\affiliation{$^4$Department of Mathematical Sciences, University of Liverpool, Liverpool, L69 7ZL, United Kingdom}

\begin{abstract}
We devise a novel duality sequence to study late-time cosmology in the heterotic ${\rm E}_8 \times {\rm E}_8$ setup of Horava and Witten with dynamical walls that are moving towards each other. Surprisingly, we find that the dimensionally reduced four-dimensional theory does not violate NEC and therefore we do not see either a bouncing or an ekpyrotic phase. Instead, our four-dimensional setup shows a transient de Sitter phase that lies well within the trans-Planckian bound. This opens up a myriad of possibilities of addressing both phenomenological and cosmological issues, and here we concentrate on one such interesting model, an axionic cosmology with temporally varying axionic coupling.
\end{abstract}

\maketitle

\newpage
\mypara{\textbf{Introduction:}}\label{sec1}
Among all the string theories, the heterotic theories and specifically the ${\rm E}_8 \times {\rm E}_8$ theory, hold a special place because of the appearance of non-abelian degrees of freedom at weak coupling and within the closed string sector. An interesting viewpoint for the ${\rm E}_8 \times {\rm E}_8$ heterotic theory from M-theory was developed by Horava and Witten \cite{horava1, horava2}. Since then many developments on the supersymmetric front appeared, but very few results were reported along the direction of the non-supersymmetric time-dependent cosmological backgrounds. One important development was the Ekpyrotic scenario \cite{ovrut1}, where the dynamical Horava-Witten walls conspire to produce a non-inflationary cosmological scenario. The construction was later realized more concretely in a four-dimensional Null Energy Condition (NEC) violating model \cite{ovrut2, ovrut3}.

In this work we provide a different cosmological model in the Horava-Witten set-up that allows for a transient late-time de Sitter phase instead. The de Sitter phase in our model is realized as an excited state over a supersymmetric minimum following earlier work of realizing such a state in type IIB theory \cite{Brahma1, Brahma2, Brahma3, Brahma4, joydeep} and in heterotic $SO(32)$ theory \cite{het32}. (A brief elaboration of this appears in section \ref{sec2}.) Two novel consequences of the construction are presented here. \textcolor{blue}{One}, a novel duality sequence that realizes a weakly coupled de Sitter excited state as a transient phase in the ${\rm E}_8 \times {\rm E}_8$ theory respecting the four-dimensional NEC, and \textcolor{blue}{two}, a novel axionic model with temporally dependent axionic decay constant satisfying the cosmological bound. The former appears in sections \ref{sec3} and \ref{sec4}, and the latter in section \ref{sec5}. We conclude in section \ref{sec6} with a brief discussion on the cosmological implications of our construction.

\mypara{\textbf{The construction of the de Sitter excited state in 
$SO(32)$ theory:}} \label{sec2}
In \cite{het32} we discussed in detail how to construct a de Sitter excited state in heterotic $SO(32)$ theory. The construction was based on earlier works for the realization of a de Sitter excited state in type IIB theory \cite{Brahma1, Brahma2, Brahma3, Brahma4}, and although both  models had their origin in the far IR limit of M-theory, there were some key differences. One such crucial distinction was with the choice of the warp factors for the internal manifold. Recall that for both  models in M-theory the underlying solitonic configurations were defined on supersymmetric warped-Minkowski backgrounds with the eleven-dimensional topology taking the form of ${\bf R}^{2, 1} \times {\cal M}_4 \times {\cal M}_2 \times {\mathbb{T}^2\over {\cal G}}$. For the type IIB case, ${\cal M}_2$ was taken to be a compact two-dimensional non-K\"ahler manifold whereas for the heterotic $SO(32)$ case it was taken to be a toroidal orbifold. The manifolds ${\cal M}_4$ for both cases were compact non-K\"ahler four-manifolds that were not required to be complex. The back-reactions of the excited states replaced the ambient supersymmetric set-up $-$ to the first approximations $-$ with time-dependent warp-factors. The existence of time-dependent warp-factors however raises different questions. For example, questions related to the existence of {\it time} and the appearance of {\it clocks}. As recently studied in \cite{wdwpaper}, the Hamiltonian constraints coming from the corresponding Wheeler-De Witt equations do allow us to define temporal evolutions in the supersymmetric Minkowski setup, although the issue of time reappears in the excited de Sitter cases. However, if we use the dual type IIA string coupling $g_s$ as {\it clocks} for both the IIB and the heterotic setup, then temporal evolutions for the excited states can be studied using the coordinate systems for the warped-Minkowski cases.  

Such a viewpoint provides a concrete way to analyze the dynamics for the type IIB and the heterotic $SO(32)$ theories. Interestingly for both cases, the excited states from M-theory take similar form as: 
\begin{align}\label{hetmet}
    ds^{2}_{11}&=g_s^{-8/3}\left(-dt^{2}+ \delta_{ij} dx^i dx^j\right) + g_s^{4/3} \delta_{ab} dw^a dw^b \\
    &+ g_s^{-2/3}{\rm H}^2(y)\left({\rm F}_1(g_s/{\rm H}) ds^2_{{\cal M}_2}+ {\rm F}_2(g_s/{\rm H}) ds^2_{{\cal M}_4}\right), \nonumber
\end{align}
where $x^i \in {\bf R}^{2, 1}$, $w^a \in {\mathbb{T}^2\over {\cal G}}$, ${\rm H}(y)$ is the remnant of the warp-factor from the supersymmetric solitonic configurations, and $g_s$ is the dual type IIA coupling such that $g_s/{\rm H}(y)$ will be a function of time in all theories which we use as clocks here. ($y^m \in {\cal M}_2 \times {\cal M}_4$ for type IIB whereas $y^m \in {\cal M}_4$ for heterotic $SO(32)$.) Note that despite the temporal dependence, no new KK modes are generated because the excited states are constructed from the fluctuation spectra of the allowed degrees of freedom at the solitonic levels that populate the far IR limit of M-theory. The difference between the type IIB and the heterotic $SO(32)$ cases now arises from the choice of the ${\rm F}_i(g_s/{\rm H})$ warp-factors. Typically we expect ${\rm F}_i(g_s/{\rm H})$ to have power law dependence on ${g_s\over {\rm H}(y)}$ whose dominant $g_s$ dependence takes the form as
${\rm F}_i(g_s/{\rm H}) = \left({g_s\over {\rm H}}\right)^{\alpha_i}$ with $\alpha_i < {2\over 3}$ so as not to jeopardize the duality sequence to the type IIB or the heterotic $SO(32)$ sides. The difference between the type IIB and the heterotic $SO(32)$ cases comes from the values assigned to $\alpha_i$. For type IIB, $\alpha_i = 0$, whereas for the heterotic $SO(32)$, $\alpha_i$'s are numbers smaller than ${2\over 3}$.

Another key difference comes from the values we assign to $g_s$ in the type IIB and the heterotic $SO(32)$ sides. For both cases we expect $g_s \to 0$ at late time \cite{Brahma1, Brahma2, Brahma3, Brahma4, desitter2}, which in fact lead to de Sitter backgrounds in flat-slicings. Moreover, the de Sitter backgrounds are transient as they are defined within the temporal domain $-{1\over \sqrt{\Lambda}} < t < 0$, where $t$ is the conformal time and $\Lambda$ is the four-dimensional cosmological constant. The dual IIA coupling $g_s \equiv g_s(t)$ is a function of the conformal time and serves as clocks for type IIB and the heterotic $SO(32)$ sides. For type IIB, $g_s = \sqrt{\Lambda}\vert t\vert$ \cite{desitter2}, and for heterotic $SO(32)$, $g_s = (\sqrt{\Lambda}\vert t \vert)^{2\over 2-\alpha_1}$ \cite{het32} with ${\rm H}(y) \equiv 1$. The latter implies that 
${\rm F}_1 =(\sqrt{\Lambda}\vert t \vert)^{2\alpha_1\over 2-\alpha_1}$.

The aforementioned way of expressing everything in terms of $g_s$ has another advantage: it helps us to quantify the dualities as {\it dynamical duality transformations}. To see this, consider the case of M-theory to type IIB. The form of $g_s$ tells us that it decreases at late time, implying that in \eqref{hetmet} the M-theory toroidal direction shrinks to zero size while all other directions increase. This leads dynamically to type IIB theory at late time after which the system does not undergo any further dualities. The IIB coupling approaches 1 \cite{desitter2}. The story in the heterotic side is more interesting. The duality sequence does not terminate after we reach type IIB because the warp-factors ${\rm F}_i$ are expressed as powers of $g_s/{\rm H}(y)$ and not constants. If we also want to keep the four-dimensional Newton's constant time-independent, then $\alpha_2 = -\alpha_1$, and this would mean that the ${\cal M}_2$ cycle starts shrinking in the IIB side (but not in the M-theory side!). Since ${\cal M}_2$ is a toroidal orbifold in M-theory, it becomes a toroidal {\it orientifold} in type IIB and therefore the dynamical duality sequence takes us to type I theory. Interestingly in type I theory the coupling now increases dynamically, taking us directly to  heterotic $SO(32)$ theory but now with a {\it weak coupling}. After which the sequence of duality transformations terminate \cite{het32}. Note that in the whole dynamical duality transformations the gauge group is broken from $SO(32)$ to $(SO(8))^4$. Interestingly, even if we arrange the duality sequence to take us via the Horava-Witten framework \cite{horava1, horava2} to heterotic ${\rm E}_8 \times {\rm E}_8$, we never recover an unbroken ${\rm E}_8 \times {\rm E}_8$. Instead the gauge group is always $(SO(8))^4$.

\mypara{\textbf{A novel dynamical duality sequence from M-theory:}}\label{sec3}
Our failure to get an unbroken ${\rm E}_8 \times {\rm E}_8$ gauge group suggests that we need to make changes to our warp-factor choices in \eqref{hetmet}. One of the simplest change is to take ${\cal M}_2$ to be an interval fibered over a circle, {\it i.e.} ${\cal M}_2 = S_{\theta_1}^1 \times {S^1_{\theta_2}\over {\cal I}_{\theta_2}}$ where ${\cal I}_{\theta_2}: \theta_2 \to - \theta_2$ with warp-factors ${\rm F}_1(g_s/{\rm H})$ and ${\rm F}_3(g_s/{\rm H})$ respectively. The distance between the two Horava-Witten walls is given by $g_s^{-1/3} \sqrt{{\rm F}_3 {\rm H}}$ and since $\alpha_i < {2\over 3}$, the walls are moving apart. This means, and as before, it is only the M-theory toroidal direction that is shrinking as $g_s^{4/3}$ and therefore 
instead of going to heterotic ${\rm E}_8 \times {\rm E}_8$ theory, we dynamically reach type IIB theory. Moreover the type IIB theory is an orientifold with orientifold seven-planes (and seven-branes) wrapped along ${\bf R}^{2, 1} \times S^1_{\theta_1} \times {\cal M}_4$. The branes and planes are points over a cylinder with topology ${\bf R}_3 \times S^1_{\theta_2}$ modded by the orientifold operation ${\cal I}_3 {\cal I}_{\theta_2} \Omega (-1)^{{\rm F}_{\rm L}}$. From M-theory we can easily infer that the base of the cylinder now shrinks as $\left({g_s\over {\rm H}}\right)^{\alpha_3}$ whereas the height increases as $\left({g_s\over {\rm H}}\right)^{-2}$, so that the overall volume of the cylinder increases. If we take ${\rm H}(y) = 1$ for simplicity (we will restore it back later), then the metric for the type IIB configuration is quite simple: it is $g_s^{-2}$ along ${\bf R}^{2, 1} \times {\bf R}_3$, and with warp-factors ${\rm F}_1, {\rm F}_3$ and ${\rm F}_2$ along $S^1_{\theta_1}$, 
$S^1_{\theta_2}$ and ${\cal M}_4$ respectively. Such a choice would have been anticipated from \cite{desitter2}, but the difference now is that in the definition ${\rm F}_i(g_s) \equiv g_s^{\alpha_i}$, $\alpha_i$ are not constants (which will be made clear in section \ref{sec4}). Moreover the global topology is quite different as is evident from the orientifold structure. When $\alpha_1 = \alpha_3$, this is similar to the heterotic $SO(32)$ case, but no novelty results from such a choice. 

We will then start by assuming that $\alpha_1$ is not equal to $\alpha_3$ although we will continue to assume that $\alpha_i < {2\over 3}$. This way the dynamical duality sequence to type IIB will not be altered at all. In fact the choice that we want to make is the following:
\begin{equation}\label{salom}
{\alpha_1\over 9} < \alpha_3 < \alpha_1 < {2\over 3}, ~~~~~ \alpha_2 < {2\over 3}, 
\end{equation}
which will make ${\rm F}_1$ to decay faster than ${\rm F}_3$. The behavior of ${\rm F}_2$ however will be arranged later on to keep the four-dimensional Newton's constant time independent. Such a choice of the internal warp-factors will imply that the type IIB configuration will now dynamically go to the type IIA theory with the radius ${\rm F}_1$ of the $S^1_{\theta_1}$ circle becoming the inverse. However because of the orientifold structure in the IIB side, the IIA configuration will inherit  a slightly more complicated orientifold of ${\bf R}_3 \times \widetilde{S}^1_{\theta_1} \times S^1_{\theta_2}$ modded out by 
$\Omega {\cal I}_{\theta_1} {\cal I}_{\theta_2} {\cal I}_3$, with $\widetilde{S}^1_{\theta_1}$ being the dual circle. This will result in orientifold six-planes (and six-branes) wrapping ${\bf R}^{2, 1} \times {\cal M}_4$.

The type IIA configuration that we got above is more complicated than naive expectation because not only the distribution of the branes and planes become dynamical, but also the IIA coupling changes from $g_s$ to ${\rm F}_1^{-1/2}$ now. Both have temporal dependence but, while the original IIA coupling from the M-theory setup showed a decrease at late time, here the behavior is inverse. However $g_s$ still continues to govern the dynamics of the system. For example, ${\bf R}_3 \times {\bf R}^{2, 1}$ still grows as $g_s^{-2}$, much like what it had in the IIB setup. The difference is that now $\widetilde{S}^1_{\theta_1}$ and ${\cal M}_4$ both grow with warp-factors proportional to ${\rm F}_1^{-1}$ and ${\rm F}_2$ respectively. (The latter is inferred from the constancy of the Newton's constant which will become clear soon.) The only direction that is shrinking in size is $S^1_{\theta_2}$ governed by the warp-factor ${\rm F}_3$. Note that the dynamical duality sequence has reversed the behavior of $S^1_{\theta_2}$.

Since $S^1_{\theta_2}$ direction appears to be shrinking, one might naively think that the system should now dynamically flow to IIB. This does not happen because the type IIA coupling is also increasing as ${\rm F}_1^{-1/2}$. Because $\alpha_1 > \alpha_3$ from \eqref{salom}, one can arrange the system in a way that $S^1_{\theta_2}$ never reaches the self-dual radius before the IIA coupling becomes much bigger than identity. The IIA model then dynamically uplifts to M-theory with the orientifold becoming an orbifold of ${\bf R}_3 \times \widetilde{S}^2_{\theta_1} \times S^1_{\theta_2}$ and the temporally varying IIA coupling turning into the new eleventh-direction with a warp-factor ${\rm F}_1^{-2/3}$.
Furthermore, the IIA branes and planes all convert to geometry which actually makes the local topology of the three-dimensional orbifold and the eleventh-circle resemble a warped combination of Atiyah-Hitchin and Taub-NUT spaces intertwined together.

\mypara{\textbf{The construction of the de Sitter excited state in 
${\rm E}_8 \times {\rm E}_8$ theory:}} \label{sec4}
Despite the complicated nature of the manifold in M-theory, the metric of the system can be {\it precisely} quantified from the dynamical duality sequence. To proceed, first note that the orbifold mentioned above can be expressed as a product of three intervals, {\it i.e.} ${{\bf R}_3 \over {\cal I}_3} \times {\widetilde{S}^1_{\theta_1} \over {\cal I}_{\theta_1}} \times {{S}^1_{\theta_2} \over {\cal I}_{\theta_2}}$. The eleventh circle, which we can call as $S^1_{11}$, is fibered over the orbifold base so as to provide the aforementioned topology of Atiyah-Hitchin and Taub-NUT spaces intertwined together. We can now blow-up two of the orbifold singularities as:
\begin{equation}\label{marcfit}
{{\bf R}_3 \over {\cal I}_3} ~ \to ~ \hat{\bf R}_3, ~~~~~~
{\widetilde{S}^1_{\theta_1} \over {\cal I}_{\theta_1}} ~ \to ~ \hat{S}^1_{\theta_1}, \end{equation}
which would keep the local metric configuration unchanged. The result of these blow-ups is to retain {\it only} the two orbifold planes coming from ${{S}^1_{\theta_2} \over {\cal I}_{\theta_2}}$ in M-theory: these are precisely the Horava-Witten walls \cite{horava1, horava2}! The fact that we could go from a system of six-branes and planes to a system of eight-branes and planes via a blow-up transformation in M-theory, is not a surprise. We have seen examples of this where a system of six-branes and planes get replaced by pure geometry via a {\it flop} transformation  in M-theory \cite{flop}. (How \eqref{marcfit} helps us to realize the full unbroken ${\rm E}_8 \times {\rm E}_8$ gauge group will be discussed in section \ref{sec5}.) In fact the presence of the two Horava-Witten walls mirrors the configuration we had right at the beginning of our dynamical duality sequence when we took 
${\cal M}_2 = S_{\theta_1}^1 \times {S^1_{\theta_2}\over {\cal I}_{\theta_2}}$. The difference now is that the non-compact manifold is ${\bf R}^{3, 1} \equiv {\bf R}^{2, 1} \times \hat{\bf R}_3$ compared to ${\bf R}^{2, 1}$ before. (Additionally, the duality direction will be 
${S^1_{\theta_2}\over {\cal I}_{\theta_2}}$, as we shall see.)
The precise metric of the excited state may be expressed as:
\begin{align}\label{hetmet2}
ds^2_{11}& = g_s^{-2} {\rm F}_1^{1/3}\Big(-dt^2  + \sum_{i,j = 1}^3 \delta_{ij} dx^i dx^j\Big)  \nonumber \\
& + {\rm F}_1^{1/3} {\rm F}_2~ ds^2_{{\cal M}_4} + {\rm F}_1^{-2/3} ds^2_{\mathbb{T}^2} + {\rm F}_1^{1/3} {\rm F}_3 ~ds^2_{\mathbb{I}}, \end{align}
where $\mathbb{T}^2 \equiv \hat{S}^1_{\theta_1} \times S^1_{11}$ and 
$\mathbb{I} \equiv {S^1_{\theta_2}\over {\cal I}_{\theta_2}}$. Thus locally the base of the seven-manifold is ${\cal M}_4 \times \mathbb{T}^2$. Of course we haven't specified the fluxes supporting the background \eqref{hetmet2}, but this is not important for what we want to do here, plus they would lead to subtle issues like flux quantization and anomaly cancellation et cetera. These questions have been addressed earlier, at least in the context of type IIB and heterotic $SO(32)$ cases in \cite{desitter2, Brahma3}, from a generic M-theory set-up so the present analysis could be thought of as forming a part of it. We will briefly discuss this later but a more detailed analysis will be presented in \cite{toappear}. Instead, in the following, we want to point out some intriguing facts about the background \eqref{hetmet2}. 

\textcolor{blue}{First}, the metric configuration \eqref{hetmet2} of the excited state solves the equations of motion coming from the Schwinger-Dyson equations. In retrospect this was expected from the fact that the dynamical duality sequence practically guarantees it provided the seed configuration in M-theory $-$ with three warp-factors ${\rm F}_1, {\rm F}_2$ and ${\rm F}_3$ $-$ solves the corresponding Schwinger-Dyson equations. This can be verified by following the EOMs laid out in \cite{desitter2, Brahma3}, and a more precise demonstration will be presented in \cite{toappear}.

\textcolor{blue}{Second}, both the four-dimensional space-time and the internal six-manifold ${\cal M}_4 \times \mathbb{T}^2$ have warp-factors that {increase} with time in the interval $-{1\over \sqrt{\Lambda}} < t < 0$. On the other hand, the distance $\rho$ between the two Horava-Witten walls, given by $\rho \equiv {\rm F}_1^{1/6} {\rm F}_3^{1/2}$,   decreases with time so that {\it the two walls are coming closer to each other}. This means at late time, when we expect $g_s \to 0$, the two walls would be right on top of each other. The four-dimensional space-time continues to expand exponentially as $g_s \to 0$ and never shows a bounce or an ekpyrotic phase. In fact as demonstrated in \cite{NEC}, the M-theory seed configuration as well as the M-theory configuration \eqref{hetmet2} {\it cannot} allow a configuration that violates NEC. Any violation of NEC in our set-up would imply that a Wilsonian EFT cannot be written down in the far IR of M-theory.

\textcolor{blue}{Third}, the observation that the two Horava-Witten walls are coming closer to each other $-$ signified by the decrease in the size of ${S^1_{\theta_2}\over {\cal I}_{\theta_2}}$ $-$ suggests that the duality sequence should not terminate in M-theory. In fact the system should now dynamically dualize to heterotic ${\rm E}_8 \times {\rm E}_8$ theory with a metric in the string frame given by:
\begin{align}\label{hetmet3}
ds^2_{10}& = g_s^{-2} \sqrt{{\rm F}_1{\rm F}_3}\Big(-dt^2  + \sum_{i,j = 1}^3 \delta_{ij} dx^i dx^j\Big) \nonumber\\
& ~~~~~~~~~~+~ {\rm F}_2\sqrt{{\rm F}_1 {\rm F}_3}~ ds^2_{{\cal M}_4} + 
\sqrt{{\rm F}_3 \over {\rm F}_1}~ ds^2_{\mathbb{T}^2}, \end{align}
where ${\cal M}_4$ and $\mathbb{T}^2$ have the same meaning as in \eqref{hetmet2}. The theory now becomes weakly coupled with a coupling constant $({\rm F}_1 {\rm F}_3^3)^{1/4}$. If we want to keep the four-dimensional Newton's constant in the heterotic side time-independent then the warp-factor ${\rm F}_2$ is defined with $\alpha_2 = -{\alpha_1 + 3\alpha_3\over 4}$. Unfortunately this makes the volume of the four-manifold ${\cal M}_4$ to {\it shrink} as $g_s^{\alpha_1 - \alpha_3}$. The 
toroidal volume then grows as $g_s^{\alpha_3 - \alpha_1}$ to keep the volume of the six-manifold time-independent. The shrinking of the four-manifold means that the duality sequence cannot terminate yet. Performing further duality transformations would either make the coupling large, or keep the total volume time-dependent, or open up avenues for even more duality transformations. None of which are attractive alternatives. However if we make $\alpha_i$, $i = 1, 3$, time-dependent as:
\begin{equation}\label{afleisA}
\alpha_i(t) = \begin{cases} ~\alpha_{io} ~~~~~~ -{1\over \sqrt{\Lambda}} < t \le -\epsilon \\
~~~ \\
~ \gamma_{io} ~~~~~~~ -\epsilon < t < 0
\end{cases}, 
\end{equation}
where $\gamma_{1o} = \gamma_{3o} = \gamma_o$ in the string frame with $\alpha_{io}$ satisfying \eqref{salom}, then at the onset of $t = \epsilon$, $\alpha_1(\epsilon) = \alpha_3(\epsilon) = \gamma_o$, and in the temporal domain $-\epsilon < t < 0$ both the internal four-manifold ${\cal M}_4$ and the torus $\mathbb{T}^2$ become time-independent.  The other parameter 
$\alpha_2(t) = -{\alpha_1(t) + 3\alpha_3(t)\over 4}$. Consequently the dynamical duality transformations would terminate. Thus in the temporal domain $-\epsilon < t < 0$, ${\rm F}_1(g_s) = {\rm F}_3(g_s)$ and the metric resembles the metric we had for the $SO(32)$ case in \cite{het32}. Once we restore back the warp-factor ${\rm H}(y)$, it is not hard to show that the string coupling and $\epsilon$ becomes:

{\footnotesize
\begin{equation}\label{larock}
{g_s\over {\rm H}(y)} = \left(\Lambda t^2\right)^{2\over 4 - \alpha_1(t) - \alpha_3(t)}, \epsilon = {1\over \sqrt{\Lambda}}\left({{\rm V}_1\over {\rm V_o}}\right)^{4 - \alpha_{1o} - \alpha_{3o} \over 2\alpha_{1o} - 2\alpha_{3o}}, \end{equation}}
giving us a transient de Sitter phase in the temporal domain $-{1\over \sqrt{\Lambda}} < t < 0$. The remaining parameters are defined as follows.
${\rm V}_o \equiv {\rm V}_o(-1/\sqrt{\Lambda}) >> {\rm V}_1 >> \alpha'^2$ is the volume of the four-manifold at $t = -{1\over \sqrt{\Lambda}}$, and ${\rm V}_1$ is the volume at $t = -\epsilon$. Plugging \eqref{larock} in \eqref{hetmet3} gives us the following metric in the ${\rm E}_8 \times {\rm E}_8$ theory in the string frame:
\begin{align}\label{readyornot}
ds^2_{10} & = {1\over \Lambda t^2} \Big(-dt^2 + \sum_{i, j = 1}^3 \delta_{ij} dx^i dx^j\Big)\nonumber\\
& + {\rm H}^4(y) \left(\Lambda t^2\right)^{\alpha_1(t) - \alpha_3(t)\over 8 - 2\alpha_1(t) - 2\alpha_3(t)} ds^2_{{\cal M}_4}\nonumber\\
& + 
\left(\Lambda t^2\right)^{\alpha_3(t) - \alpha_1(t)\over 4 - \alpha_1(t) - \alpha_3(t)} ds^2_{\mathbb{T}^2}, \end{align}
which is a transient de Sitter excited state defined in the interval $-{1\over \sqrt{\Lambda}} < t < 0$. The $\alpha_i(t)$ parameters, namely
$\alpha_1(t)$ and $\alpha_3(t)$, take the values as in \eqref{afleisA}, which implies that the internal six-manifold becomes completely time-independent in the temporal domain $-\epsilon < t < 0$. 

\mypara{\textbf{Axions and axionic couplings in heterotic ${\rm E}_8 \times {\rm E}_8$ theory:}}\label{sec5} The dynamical motion of the Horava-Witten walls, leading to the transient de Sitter phase as in \eqref{readyornot}, now directs us to analyze many phenomenological issues in a cosmological set-up. Here we will only address the story with the axions and the axionic couplings. However the study of axions is done mostly using Einstein frame metric which may be derived from \eqref{readyornot} by knowing the heterotic coupling. As seen above, the heterotic theory is weakly coupled with a coupling constant given by:

{\footnotesize
\begin{equation}\label{samweavin}
g_{\rm het} = {\rm H}^2(y) \left({g_s\over {\rm H}(y)}\right)^{\alpha_1 + 3\alpha_3\over 4}, {g_s\over {\rm H}(y)} = \left(\Lambda t^2 \right)^{8 \over 16 - 3\alpha_1(t) -\alpha_3(t)}, \end{equation}}
where we do not use \eqref{larock} because they were derived using string frame metric. In the Einstein frame metric, the heterotic coupling as given in \eqref{samweavin} is arranged to get a transient de Sitter excited state that keeps the four-dimensional Newton's constant time-independent. As such this makes $\alpha_2(t) = -{\alpha_1(t) + 3\alpha_3(t) \over 16}$, and the Einstein frame metric takes the following form:
\begin{align}\label{trotttagre}
ds^2_{10} & = {1\over \Lambda t^2 {\rm H}(y)} \Big(-dt^2 + \sum_{i, j = 1}^3 \delta_{ij} dx^i dx^j\Big)\nonumber\\
& + {\rm H}^3(y) \left(\Lambda t^2 \right)^{5\alpha_1(t) - \alpha_3(t)\over 32 - 6\alpha_1(t) - 2\alpha_3(t)} ds^2_{{\cal M}_4}\nonumber\\
& + {\rm H}^{-1}(y)
\left(\Lambda t^2 \right)^{\alpha_3(t) - 5\alpha_1(t)\over 16 - 3\alpha_1(t) - \alpha_3(t)} ds^2_{\mathbb{T}^2}, \end{align}
which may be compared to \eqref{readyornot}. Observe that we cannot get de Sitter transient state in both frames simultaneously but we can come close ({\it i.e.} upto the overall warp-factor ${\rm H}(y)$) if $\alpha_i = 0$. Since we are interested in $\alpha_i \ne 0$, we will ignore such possibilities. Note that if $\gamma_{1o} = \gamma_o$, then $\gamma_{3o} = 5\gamma_o$ in \eqref{afleisA} and $\alpha_{io}$ satisfies \eqref{salom} in the Einstein frame.

Let us also make a few consistency checks. \textcolor{blue}{One}, the choice of the coupling constants \eqref{samweavin} in the heterotic side suggests that the quantity $n \equiv {16\over 3\alpha_1(t) + \alpha_3(t) - 16}$ should be 
away from the regime $-1 < n < 0$ for the validity of EFT \cite{NEC, Brahma3}. This is clearly the case here and even for $\alpha_1(t) = \alpha_3(t) = 0$ suggesting no violation of the four-dimensional NEC. 
\textcolor{blue}{Two}, the blow-ups that we discussed in \eqref{marcfit}
lead to two orientifold eight-planes with $-16$ units of RR charges each. As such one would expect to cancel them by putting 16 D8-branes on top of each of the two O8-planes. At the solitonic level this breaks supersymmetry, so instead we can separate one D8-brane from each of the two sets of eight-branes to create type I$'$ half D-particles \cite{gorbor}. These half D-particles create extra generators that enhance each $SO(14) \times U(1)$ to ${\rm E}_8$. In other words:
\begin{equation}
{\bf 248} = {\bf 1}_0 \oplus {\bf 91}_0 \oplus {\bf 14}_{+1} \oplus {\bf 14}_{-1} \oplus {\bf 64}_{+{1\over 2}} \oplus {\bf 64}_{-{1\over 2}}, \end{equation}
where the subscripts denote the $U(1)$ charges. The enhancement being a non-perturbative process, all the irreps other than ${\bf 1}$ and ${\bf 91}$ have to be realized non-perturbatively. Thus this is the way we can get unbroken ${\rm E}_8 \times {\rm E}_8$ in our setup. \textcolor{blue}{Three}, the duality sequence that we performed are in a non-supersymmetric setup so one might be worried if this is at all possible. The answer is that, at the solitonic level the dualities are {\it always} over supersymmetric backgrounds. At every stage of the dualities, the corresponding non-supersymmetric (or more appropriately spontaneously broken supersymmetric) backgrounds are constructed over the supersymmetric backgrounds using the Glauber-Sudarshan states, much like how we did it in \cite{Brahma1, Brahma2, Brahma3, Brahma4}. Moreover, at every stage the non-supersymmetric backgrounds have to satisfy the corresponding Schwinger-Dyson equations (see for example \cite{joydeep}). The dynamical duality sequence provides the {\it dominant} contributions in terms of powers of $g_s$ with the subdominant ones coming from the aforementioned Schwinger-Dyson equations \cite{toappear}. \textcolor{blue}{Four}, the fluxes which are important in the construction of the axions can also be traced right from the start in M-theory with ${\cal M}_2 \equiv S^1_{\theta_1} \times {S^1_{\theta_2}\over {\cal I}_{\theta_2}}$. At the solitonic level the background has 16 real supercharges and the three-form field ${\bf C}$ is defined with one leg along the ${S^1_{\theta_2}\over {\cal I}_{\theta_2}}$ direction so as not to be projected out by the orbifold action. The subtleties however are the intermediate orientifold actions and the eventual blow-ups \eqref{marcfit}. In the end what survives are the flux components ${\bf C}_{\theta_2 {\rm MN}}$ with $({\rm M, N}) \in {\bf R}^{3, 1}$ or $({\rm M, N}) \in {\cal M}_4 \times \mathbb{T}^2$. The former would lead to {\it model-independent} axions and the latter to the {\it model-dependent} axions \cite{Svrcek:2006yi}. The model-independent axion (which is actually a misnomer as shown in \cite{hassan}) would have interfered with the de Sitter isometries, but since we are dealing with transient de Sitter phase this is not much of a concern.

 With the aforementioned computations and checks at hand, we are ready to deal with the cosmological implication of the axion field. As mentioned earlier, we will be in the Einstein frame where the four-dimensional Newton's constant is related to ${\rm M}_p, {\rm M}_s, {\rm H}(y)$ and the unwarped volume of the internal six-manifold as:
 \begin{equation}\label{samsky}
 {1\over 8\pi {\rm G}_{\rm N}} = {\rm M}_p^2 = 4\pi {\rm M}_s^8 \int d^6 y~
 {\rm H}^4(y) \sqrt{{\rm det}~g_{{\cal M}_4}\cdot {\rm det}~g_{\mathbb{T}^2}}, \end{equation}
 where we have used the ten-dimensional heterotic metric \eqref{trotttagre}. The rest of the computation {\it almost} follows the standard procedure. For the model independent axions we will identify them from the heterotic three-form ${\bf H}_3$ along ${\bf R}^{3, 1}$ directions. (The model dependent axions can also be studied in our model $-$ by taking the antisymmetric tensors tangent to the internal directions and viewing them as four-dimensional scalars using the harmonic two-forms on ${\cal M}_4 \times \mathbb{T}^2$ $-$ but we will not do so here.) The relevant part of the dimensionally reduced ten-dimensional heterotic action in the Einstein frame is:
 \begin{align}\label{sirsays}
 \mathbb{S}_4 & = -{{\rm M}_p^2\over 4}\int \beta(t)~ {\bf H}_3 \wedge \ast_4 {\bf H}_3\nonumber\\
 & + \int a\left[d{\bf H}_3 - {\alpha'\over 4}\Big({\rm tr}~{\bf R}_4 \wedge {\bf R}_4 - {1\over 30}{\rm Tr}~{\bf F} \wedge {\bf F}\Big)\right], \end{align}
 where $\beta(t)$ is a parameter that depends on the specifics of the dimensional reduction, {\it i.e.} on the unwarped volumes of the sub-manifolds ${\cal M}_4$ and $\mathbb{T}^2$ from \eqref{trotttagre}, as well as the heterotic dilaton from \eqref{samweavin}. This develops an explicit time dependence because of the time dependence of the dilaton, and can be expressed in the following way:
 \begin{equation}\label{maamey}
 \beta(t) = \left(\Lambda t^2\right)^{2\alpha_1 + 6\alpha_3\over 3\alpha_1 + \alpha_3 - 16} , \end{equation}
where $\alpha_i = \alpha_i(t)$ is defined by \eqref{afleisA} which, as we shall see, will be modified further. The other parameter $a$, appearing as a Lagrange multiplier in \eqref{sirsays}, will be related to our axionic field. We can now integrate out ${\bf H}_3$ and express \eqref{sirsays} completely in terms of the field $a$ in the following way (see also section 2 of \cite{hassan}):
\begin{align}\label{milenjon}
\mathbb{S}_4 & = {2\over {\rm M}_p^2} \int {d^4x\over \beta(t)} \Big(-{1\over 2}\partial_\mu a \partial^\mu a\Big)\nonumber\\
& - \int {a\alpha'\over 4}\Big({\rm tr}~{\bf R}_4 \wedge {\bf R}_4 - {1\over 30}{\rm Tr}~{\bf F} \wedge {\bf F}\Big), \end{align}
where note that compared to say \cite{hassan} we have the $\beta(t)$ factor inside the integral. Unfortunately this means that we cannot simply use the standard manipulation to absorb the $\beta(t)$ factor in the definition of $a$. To facilitate the computations let us then define the field $a$ in the following suggestive way:
\begin{align}\label{milenrmey}
a({\bf x}, t) & = \int_\alpha^t dt'\sqrt{\beta(t')} {\partial\over \partial t'}\left({{\rm A}({\bf x}, t')\over \sqrt{\beta(t')}}\right)\\
& = {\rm A}({\bf x}, t) - {\rm A}({\bf x}, \alpha) - {1\over 2} \int_\alpha^t dt' ~{\partial\beta(t')\over \partial t'} \cdot {{\rm A}({\bf x}, t')\over \beta(t')}, \nonumber \end{align}
where ${\rm A}({\bf x}, t)$ would be our axionic field, $\alpha^{-2} = \Lambda$ with $\alpha < 0$, ${\bf x} \in {\bf R}^3$, and $\beta(t)$ is defined in \eqref{maamey}. One of the advantage of defining the axionic field ${\rm A}({\bf x}, t)$ in the aforementioned way is to observe that the temporal and spatial derivatives of $a$ take the following form (${\rm A}({\bf x}, \alpha)\equiv 0$):

{\footnotesize
\begin{align}\label{jbrewsterj}
& {\partial a({\bf x}, t)\over \partial t}  = \sqrt{\beta(t)} {\partial\over \partial t}\left({{\rm A}({\bf x}, t)\over \sqrt{\beta(t)}}\right)\\
& {{\bf \nabla} a({\bf x}, t)} = \sqrt{\beta(t)} ~{\bf \nabla}\left({{\rm A}({\bf x}, t)\over \sqrt{\beta(t)}}\right) - {1\over 2} \int_\alpha^t dt'{\dot{\beta}(t')\over \beta(t')}
~ {\bf \nabla}{\rm A}({\bf x}, t'), \nonumber \end{align}}
which almost reproduces what we want except that the gradient has an extra contribution. Interestingly both the expression for the gradient in \eqref{jbrewsterj} as well as for ${\rm A}({\bf x}, t)$ in \eqref{milenrmey} have small non-local contributions coming from the integration over the temporal coordinate $t$. Plugging \eqref{jbrewsterj}, \eqref{milenrmey} and \eqref{maamey} in \eqref{milenjon}
we get the requisite action for the axion ${\rm A}({\bf x}, t)$ in the standard form $-$ except with additional interactions coming from the shifts in \eqref{jbrewsterj} and \eqref{milenrmey} $-$ with the axion coupling $f_a$ taking the following functional value: 

{\scriptsize
\begin{align}\label{salamthai}
f_a(t) = \sqrt{2\over \beta(t)}~{{\rm M}_s^2\over {\rm M}_p} & = {1\over 2\pi}
\left(\Lambda t^2\right)^{\alpha_1(t) + 3\alpha_3(t) \over 16 - 3\alpha_1(t) - \alpha_3(t)}\\
& \times \Bigg({2\pi {\rm M}_s^{-4} \over \int d^6 y ~{\rm H}^4(y) \sqrt{{\rm det}~g_{{\cal M}_4}(y)\cdot {\rm det}~g_{\mathbb{T}^2}(y)}}\Bigg)^{1/2},\nonumber \end{align}}
where the temporal modulating factor in the axion coupling starting at $ t = -{1\over \sqrt{\Lambda}}$ shows a general decrease towards smaller values as $\alpha_i(t)$, for $i = 1, 3$, appearing in \eqref{salamthai} are mostly constant over a wide temporal range from \eqref{afleisA}. However since the value of $\alpha_i(t)$ does change to $\gamma_{io}$ in the temporal domain $-\epsilon < t < 0$, there is a possibility of an intermediate maxima unless we demand the following lower bound on $\gamma_o$ (recall that $\gamma_{1o} = \gamma_o, \gamma_{3o} = 5\gamma_o$ in \eqref{afleisA}):
\begin{equation}\label{ivbres}
\gamma_o \ge {2\alpha_{1o} + 6\alpha_{3o}\over 32 - 5\alpha_{1o} + \alpha_{3o}}, \end{equation}
to allow for a monotonic decrease of $f_a(t)$. Such a choice turns out to be easily arranged \cite{toappear} and therefore we expect $f_a(t)$ to 
 finally approach zero as we go to late time (recall that our de Sitter excited state is defined with a flat-slicing in the temporal domain $-{1\over \sqrt{\Lambda}} < t < 0$, so $\Lambda t^2$ monotonically decreases). This is problematic, but let us first analyze the simpler case associated with \eqref{salamthai} before commenting on the problem. When $\alpha_i(t) = 0$ and ${\rm H}(y) = 1$, which is the case with no warp-factors, the integral over $y^m$ ($m \in {\cal M}_4 \times \mathbb{T}^2$) simply gives us ${\rm M}_s^2{\rm M}_p^{-1}$ from \eqref{samsky}. For this case ${{\rm M}_s\over {\rm M}_p} \approx {1\over 18}$ and $f_a$ in \eqref{salamthai} becomes 
$f_a \approx 10^{16} {\rm GeV}$, a value too large to be acceptable \cite{Svrcek:2006yi}. Switching on ${\rm H}(y)$ and $\alpha_i$, the situation improves. At time $t > -{1\over \sqrt{\Lambda}}$, with a roughly constant value  ${\rm H}(y) \ge 100$, the decrease in $f_a$ can easily make it to lie within the allowed range of $10^9 ~{\rm GeV} < f_a < 10^{12}~ {\rm GeV}$ \cite{hassan}. However, as mentioned above, the puzzle is at late time when $t \to 0$. Since $\alpha_i(t)$ approaches $\gamma_{io}$ as shown in \eqref{afleisA}, $f_a(t) \to 0$ as $t \to 0$.  A simple way out of this would be to take $\gamma_o = 0$ which, clearly from \eqref{ivbres}, does not allow for a monotonically decreasing $f_a(t)$ but makes $f_a(t)$ to jump. This is yet again not acceptable. An alternative way is to take $\gamma_o$ to actually become dynamical as:
\begin{equation}\label{heigelK}
\gamma_o(t) = {2c_o\over c_o + 2\vert\log(\Lambda t^2)\vert},  ~~~~ -\epsilon < t < 0, \end{equation}
with constant $c_o \equiv {\alpha_{1o} + 3\alpha_{3o}\over 16 - 3\alpha_{1o} - \alpha_{3o}}~\vert\log(\Lambda\epsilon^2)\vert$ so that the temporal part of $f_a(t)$ becomes $e^{-c_o}$ in the temporal domain $-\epsilon < t < 0$. Happily, this again makes $f_a(t)$ to lie within the allowed range and consequently terminates the motion of the walls in the range $-\epsilon < t < 0$.
A more general solution exists, whose analysis unfortunately takes us beyond the scope of this paper. We will elaborate on this in \cite{toappear}. 

\mypara{\textbf{Discussion:}} \label{sec6} In this paper we presented the construction of a transient de Sitter phase in heterotic ${\rm E}_8 \times {\rm E}_8$ theory that allows us to study a novel axionic theory with time-dependent axionic coupling. Both the transient de Sitter phase and the axionic coupling lie well within their respective cosmological bounds. However we haven't discussed the broader cosmological implications of our model, for example the connection to weak gravity conjecture \cite{vafa} et cetera. This and other related topics will be discussed elsewhere.

%\vspace{2mm}

\noindent{\bf Acknowledgments:} We would like to thank R. Brandenberger, E. Hardy and J. Smirnov for helpful discussions. The work of SB 
supported in part by the Higgs Fellowship and by the STFC Consolidated Grant
“Particle Physics at the Higgs Centre”. The work of KD and BK is supported in part by a Discovery Grant from NSERC. The work of AM is supported in part by the Prime Minister’s Research Fellowship provided by the Ministry of Education, Government of India. The work of RT is supported by STFC Consolidated Grant ST/X000699/1.

\bibliographystyle{apsrev4-1}
\bibliography{references.bib}
\end{document}